\documentclass[12pt]{article}
\usepackage{graphicx}
\usepackage{amsmath}
\usepackage{amssymb}
\usepackage{amsfonts}
\usepackage{caption}
\usepackage{mathtools}
\usepackage{color}
\usepackage[utf8]{inputenc}
\usepackage[english]{babel}
\usepackage[table]{xcolor}
\usepackage{setspace}
\usepackage{pdfpages}
\allowdisplaybreaks

\usepackage{array}
\usepackage{cellspace}
\usepackage[left=2.5cm,right=2.5cm, top=3cm, bottom=3cm]{geometry}
\newcolumntype{L}[1]{>{\raggedright\arraybackslash} m{#1} }
\usepackage[breaklinks=true,colorlinks=true,linkcolor=blue,urlcolor=blue,citecolor=blue]{hyperref}

\begin{document}
\title{\Huge The extension of Schwarzschild line element to include uniformly accelerated mass}

  \author{\text{Ranchhaigiri Brahma$^{1\ast}$ } and \text{A.K. Sen$^{1\dagger}$  }\\
  \\ \textsl{$^1$Department of Physics, Assam University, Silchar-788011, India}\\ $^{\ast}$ Email: \href{mailto: rbrahma084@gmail.com}{\textit{ rbrahma084@gmail.com}}\\
  $^{\dagger}$ Email: \href{mailto: asokesen@yahoo.com}{\textit{ asokesen@yahoo.com}}}
\date{  }
\maketitle
\onehalfspacing
\abstract{In this paper we analyze the spacetime geometry due to a Schwarzschild object having uniform accelerated  motion. In the beginning, we investigate the gravitational field due to a uniformly moving Schwarzschild object and obtain the spacetime line element for such an object. After analyzing the necessary limiting conditions, the obtained line element is found to be consistent. Next, we extend our work to a uniformly accelerated Schwarzschild object. In that case, we obtain the spacetime line element for both when the acceleration is along the $X-$direction and when the acceleration is in an arbitrary direction on the $XY$ plane (in the  later  case  $X$  may  be  some  third reference  direction,  which may  be useful  for  future work). The limiting conditions of those line elements have been examined. Such work will have relevant applications in astrophysics when we calculate the geodesic equations of test particles in the gravitational field of objects having uniform accelerated motion.}

\section{Introduction} \label{sec:intro}
In general relativity the gravitational field of a static or a stationary object and the dynamics of the particles in such a field have been studied extensively in the past (e.g. \cite{einsteina, landau, weinberg, misner, bibchnadra} etc.). But not much work has been done where the line element due to a moving gravitational mass has been calculated without any PPN approximation. So, it remains as one of the fascinating and hard subjects in the field of theoretical analysis. Although the study of the gravitational field due to moving objects is quite complicated, in a few cases e.g. \cite{1971GReGr...2..303A, 1993ApJ...415..459P, PhysRevD.65.064025, 2003ApJ...586..731M, PhysRevD.75.062002, 10.1111/j.1365-2966.2009.15387.x} the authors adopted various procedures to investigate the gravitational field of moving objects and suggested the dynamics of the particles in those fields.
One of the interesting methods that some authors have used is the method of coordinate transformation to obtain the metric for moving objects. Guan-Sheng and Wen-Bin \cite{2014CoTPh..61..270H} obtained the spacetime line element for a uniformly moving point mass by performing the Lorentz transformation. Using the same approach, the gravitational field for moving Reissner-Nordstr\"om (R-N) black hole and moving Kerr black hole has been studied in \cite{He_2014} and \cite{2014IJMPD..2350031H} respectively. On the other hand, Paithankar and Kolekar \cite{PhysRevD.99.064012} discussed about a Schwarzschild body, in the
Minkowski-Rindler setting such that the black hole is at a large distance away from the Rindler observer. They also mentioned that, in this situation the spacetime geometry in the local neighborhood of the Rindler trajectory will be dominantly flat with small perturbations to the background flat metric due to the presence of the black hole. Later, they investigated the trajectories of linear uniform acceleration in the background of Schwarzschild geometry and compared their results with that of the $C$-metric. The $C$-metric is an exact solution of the Einstein field equation, which represents a pair of black holes with equal mass accelerated away from each other due to the presence of conical singularities \cite{GRIFFITHS_2006a, Griffiths_2006}. Studies on light-like  geodesics and time-like geodesics in $C$-metric spacetime can be found in \cite{Frost_2021} and \cite{PhysRevD.106.044037}, respectively.  However, to our knowledge, no work has been done using the coordinate transformation approach to investigate the gravitational field due to uniformly accelerated or arbitrarily accelerated gravitating objects except a few, as like in the case of uniformly moving black holes \cite{1971GReGr...2..303A}. Therefore, it is necessary to study the gravitational field due to an accelerated object using the coordinate transformation method.\\\\
The objective of the current work is to investigate the gravitational field, that is, the spacetime geometry due to the uniformly accelerated Schwarzschild object. So, at first we started our work by investigating the gravitational field due to a uniformly moving Schwarzschild object and obtained the spacetime line element by using the method of coordinate transformation i.e. using the Lorentz transformation.  In the case of an accelerated frame, the general transformation relations are presented in \cite{nelson1987generalized,1994JMP....35.6224N}. The coordinate transformations for an accelerated frame with constant proper acceleration is known as the Rindler transformation, and the consequences of such coordinate transformations have been studied in  \cite{PhysRev.119.2082, rindler1966kruskal, David1997, Yi}. 
In the Rindler transformation, the initial velocity is considered as zero. However, the coordinate transformation for the accelerated frame with non-zero initial velocity has been discussed in \cite{Yi} and we adopted this procedure here to obtain the transformation relations in order to study the metric for an accelerated Schwarzschild object. Subsequently, we also calculated the Rindler transformation having acceleration along an arbitrary direction on the $XY$ plane. Here  we note,  in the  later  case  $X$  may  be  some  third reference  direction,  which may  be useful  for  future work.  So  we  are  essentially  decomposing  the acceleration into  two  arbitrarily  oriented  orthogonal directions. This decomposition we have done, keeping in mind the scope for future work. For example, by considering two simple harmonic motions in two orthogonal directions separated by a phase angle of the $\pi/2$ radians, we can also generate circular / elliptical motions. We then construct the spacetime line element due to a Schwarzschild object uniformly accelerated along some arbitrary direction on the $XY$ plane. Then we examined the obtained line elements under various limiting conditions. Such work will have relevant applications in astrophysics when we calculate the geodesic equations of test particles in the gravitational field of objects having uniform acceleration.\\\\
We organize our paper as follows. In {\bf Section 2} we discuss the isotropic form of the Schwarzschild line element. Then, we obtain a far-field approximation of the Schwarzschild line element in the Cartesian coordinates. In {\bf Section 3} we investigate the gravitational field due to a uniformly moving Schwarzschild object. Consequently, we obtained the line element for such an object in terms of spatial Cartesian coordinates. Similarly in {\bf Section 4} we calculated the spacetime line element due to a uniformly accelerated Schwarzschild object along $X-$direction and also along some arbitrary direction on the $XY$ plane. Finally, in {\bf Section 5} we discuss the results obtained in the above calculations and conclude our findings.

\section{Schwarzschild line element under far-field approximation using isotropic cartesian coordinate system}
The gravitational field and hence the spacetime geometry due to a static and spherically symmetric object can be described with the well known Schwarzschild line element and we call such object the Schwarzschild object \cite{schwarzschild}. The Schwarzschild line element can be written as follows:
\begin{align}
    ds^2&=\left(1-\frac{2m}{r}\right)c^2dt^2-\left(1-\frac{2m}{r}\right)^{-1}dr^2-r^2d\theta^2-r^2\sin^2\theta\,d\phi^2\label{eq:x1a}
\end{align}
where $m$ is the mass parameter which is equal to half of the Schwarzschild radius $r_g=2GM/c^2$, $c$ is the speed of light in vacuum, $G$ is the gravitational constant, and $M$ is the mass of the object. The line element \eqref{eq:x1a} can be written in isotropic form by assuming a radial function $\rho$ which is related to the radial coordinate $r$ of the spherical polar coordinate as follows \cite{landau, weinberg}:
\begin{align}
     r&=\left(1+\frac{m}{2\rho}\right)^2\rho\label{x3}
\end{align}
So using the relation \eqref{x3}, we can obtain the isotropic form of the Schwarzschild line element as below \cite{landau, weinberg}:
\begin{align}
    ds^2&=\left(\frac{1-\frac{m}{2\rho}}{1+\frac{m}{2\rho}}\right)^2c^2dt^2-\left(1+\frac{m}{2\rho}\right)^4\Big\{d\rho^2+\rho^2d\theta^2+\rho^2\sin^2\theta\,d\phi^2\Big\}\label{eq:x1}
\end{align}
where $\rho$, $\theta$, $\phi$ are the isotropic spherical coordinates. Now we introduce isotropic Cartesian coordinates $x$, $y$, $z$ and further we apply far-field approximation, i.e. $\rho>>m$. Therefore we can approximately write from equation \eqref{eq:x1} as follows \cite{landau}:
\begin{align}
    ds^2=\left(1-\frac{2m}{\rho}\right)c^2dt^2-\left(1+\frac{2m}{\rho}\right)\Big\{dx^2+dy^2+dz^2\Big\}\label{neq2}
\end{align}
Again, from equation \eqref{x3} expanding $r/\rho$ in the power of $m/\rho$ we obtain as:
\begin{align}
    \frac{r}{\rho}&\approx 1+\frac{m}{\rho}+\frac{m^2}{4\rho^2}
\end{align}
Since $\rho>> m$, so neglecting the terms higher than $m/\rho$, we can obtain as:
\begin{align}
 \rho&\approx r-m\label{x6}
\end{align}
Then substituting from equation \eqref{x6} into equation \eqref{neq2} we can obtain a far-field approximated Schwarzschild line element as: 
\begin{align}
    ds^2=\left(1-\frac{2m}{r-m}\right)c^2dt^2-\left(1+\frac{2m}{r-m}\right)\Big\{dx^2+dy^2+dz^2\Big\}\label{x7}
\end{align}
where $r=\sqrt{x^2+y^2+z^2}$. \\\\
We use the line element \eqref{x7} to study the gravitational field due to a Schwarzschild object undergoing uniform motion and accelerated motion. 

\section{\label{sec:level2}Metric for a uniformly moving Schwarzschild object }
Now, we consider that the Schwarzschild object is moving with uniform velocity along the $X-$direction. Then for the sake of convenience in studying the gravitational field, we rewrite the line element \eqref{x7} in terms of the usual primed coordinate frame $K'$ used as a moving frame in special relativity as follows:
\begin{align}
    ds^2=\left(1-\frac{2m}{r'-m}\right)c^2dt'^2-\left(1+\frac{2m}{r'-m}\right)\Big\{dx'^2+dy'^2+dz'^2\Big\}\label{eq:x3}
\end{align}
where $r'=\sqrt{x'^2+y'^2+z'^2}$.
Then we consider that the coordinate frame $K'(ct', x', y', z')$ is moving with respect to the observer's rest frame $K(ct, x, y, z)$ with uniform velocity $v_x$ in $X-$direction. Additionally, we also consider that for the frames $K$ and $K'$ the respective $X$ and $X'$ axes are coinciding and other coordinate axes $(Y, Z)$ and $(Y', Z')$ are parallel to each other. In curved spacetime (viz Schwarzschild) the relation between two such frames cannot be directly established by using the standard Lorentz transformation relations.  It is  because  Lorentz transformation  uses specific
properties of the Minkowski spacetime such as the parallel transport of vectors
allowed by its very particular affine structure, which the Schwarzschild geometry does not allow. However, in our present work, we considered a far-field condition where the spacetime geometry is close to flat, that can be described using the line element in equation \eqref{eq:x3}. So in this context, we can use the Lorentz transformation relations written in terms of Cartesian coordinates as follows:
\begin{equation}\label{eq:10}
    \begin{aligned}
    ct&=\gamma\left(ct'+\frac{v_x}{c} \cdot x'\right),\hspace{0.5cm}
    x=\gamma\left(x'+v_x\,t'\right), \hspace{0.5cm}y&=y'\,\,,\hspace{0.5cm}
     z=z'
\end{aligned}
\end{equation}
where $\gamma=\left(1-v_x^2/c^2\right)^{-1/2}$. Furthermore, the inverse relation of equation \eqref{eq:10} can be written as:
\begin{equation}\label{eq:10a}
    \begin{aligned}
    ct'&=\gamma\left(ct-\frac{v_x}{c} \cdot x\right),\hspace{0.5cm}
    x'=\gamma\left(x-v_x\,t\right), \hspace{0.5cm}y'&=y\,\,,\hspace{0.5cm}
     z'=z
\end{aligned}
\end{equation}
Therefore, using \eqref{eq:10a} we can write:
\begin{align}
    cdt'&=\gamma\left(c dt- \frac{v_x}{c}dx\right),\,\,\,
    c^2 dt'^2=\gamma^2\left[ c^2(dt)^2+ \frac{v_x^2}{c^2}(dx)^2-2\frac{v_x}{c}dxcdt\right] \\
     dx'&=\gamma\left(- v_x dt+ dx\right)\text{\,\,\,\,or\,\,\,\,}
    dx'^2=\gamma^2\left[ v_x^2 (dt)^2+ (dx)^2-2v_xdxdt\right]\\
    dy'&=dy\,\,\,\Rightarrow dy'^2=(dy)^2\text{\,\,\,\,and\,\,\,\,}
    dz'=dz\,\,\, \Rightarrow dz'^2=(dz)^2\label{eq:L11}
\end{align}
Now following the similar approach used in \cite{1971GReGr...2..303A, 1993ApJ...415..459P, 2014CoTPh..61..270H, He_2014} we obtain the line element by putting \eqref{eq:10a}--\eqref{eq:L11} in equation \eqref{eq:x3} and simplifying as below:
\begin{align}
    ds^2&= \Bigg[1-\frac{2 m\gamma ^2}{\mathcal{R}}\Big\{1+\frac{v_x^2}{c^2}\Big\}\Bigg]c^2dt^2- \Bigg[1+\frac{2 m\gamma ^2}{\mathcal{R}}\Big\{1+\frac{v_x^2}{c^2}\Big\}\Bigg]dx^2\nonumber\\
    &-\Big[1+\frac{2 m}{\mathcal{R}}\Big]\Big\{dy^2+dz^2\Big\}+\frac{8 \gamma ^2 m v_x}{\mathcal{R}}\,dx dt \label{eq:x10}
\end{align}
where $\mathcal{R}=\sqrt{\gamma ^2 \left(x-v_x\,t\right)^2+y^2+z^2}-m$. 
The line element \eqref{eq:x10} represents the gravitational field of a uniformly moving Schwarzschild object. Below we analyze this line element at different limiting conditions:
\begin{enumerate}
    \item[(a)]{\bf Case-I:} If $v_x=0$ then $\gamma=1$ and hence we obtain from \eqref{eq:x10} as:

\begin{align}
    ds^2&=\left[1-\frac{2m}{r-m}\right]c^2dt^2-\left[1+\frac{2m}{r-m}\right]\Big\{dx^2+dy^2+dz^2\Big\}\label{x11q}
\end{align}
which is exactly similar to the Schwarzschild line element in Cartesian coordinates as shown in equation \eqref{eq:x3} as $ct=ct'$, $x=x'$, $y=y'$, $z=z'$ and $r=r'$ for $v_x=0$.
\item[(b)] {\bf Case-II:} If the mass is zero then the mass parameter $m=0$ and hence from equation \eqref{eq:x10} we obtain as:
\begin{align}
    ds^2&=c^2dt^2-dx^2-dy^2-dz^2
\end{align}
Thus the line element obtained in \eqref{eq:x10} reduces to the Minkowski line element (the line element for flat spacetime) in the absence of mass.
\item[(c)] Case-III: Following a similar approach used in \cite{1971GReGr...2..303A, Cristofoli_2020} we calculate a spacetime metric from equation \eqref{eq:x10} in which the velocity of the object is close to the speed of light i.e. $v_x\rightarrow c$. Here using the relativistic mass $M_r$ and energy $E$ relation: $E=M_rc^2$, we write $E=\frac{Mc^2}{\sqrt{1-\frac{v_x^2}{c^2}}}$ and hence obtain $m=\frac{GM}{c^2}=\frac{GE}{c^4}\sqrt{1-\frac{v_x^2}{c^2}}$, where $M$ is the rest mass and $G$ is the gravitational constant. Then substituting this expression for $m$ into equation \eqref{eq:x10} we obtain as:
\begin{align}
   ds^2
    &= \Bigg[1-\frac{2 GE}{c^4\mathcal{R}_0}\Big\{1+\frac{v_x^2}{c^2}\Big\}\Bigg]c^2dt^2- \Bigg[1+\frac{2GE }{c^4\mathcal{R}_0}\Big\{1+\frac{v_x^2}{c^2}\Big\}\Bigg]dx^2\nonumber\\
    &-\Big[1+\frac{2 GE}{c^4\mathcal{R}_0 }\Big\{1-\frac{v_x^2}{c^2}\Big\}\Big]\Big\{dy^2+dz^2\Big\}+\frac{8 GE  v_x}{c^4\mathcal{R}_0}dx dt\label{x17}
\end{align}
where $\mathcal{R}_0=\Big\{ \left(x-v_x\,t\right)^2+\Big\{1-\frac{v_x^2}{c^2}\Big\}(y^2+z^2)\Big\}^{1/2}-\frac{GE}{c^4}\Big\{1-\frac{v_x^2}{c^2}\Big\}$. In the limit of $v_x\rightarrow c$, the above line element \eqref{x17} can be written as:
\begin{align}
    ds^2& =c^2dt^2-dx^2-dy^2-dz^2-\frac{4GE}{c^4\ (x-c t)}\Big\{cdt-dx\Big\}^2\label{qx11}
\end{align}
which is exactly same as the Aichelburg-Sexl metric\footnote{In the original derivation, Aichelburg and Sexl \cite{1971GReGr...2..303A} considered $c=G=1$ and $E=p$ as a constant.} derived in \cite{1971GReGr...2..303A, Cristofoli_2020}. The difference is only in the choice of the reference frame to represent the equation \eqref{qx11} where the original Aicherburg-Sexl metric is written in terms of the primed coordinate frame.  
Therefore, we can state that the line element \eqref{eq:x10} reduces to the Aicherburg-Sexl metric under the condition $v_x\rightarrow c$.
 
\end{enumerate}
In our present study, we observed that the currently obtained line element \eqref{eq:x10} satisfies the necessary limiting conditions, and hence this line element can describe the spacetime geometry due to a uniformly moving Schwarzschild object. Such line element for a uniformly moving Schwarzschild object was also obtained by Guan-Sheng and Wen-Bin \cite{2014CoTPh..61..270H} in harmonic form. The line element obtained there also recovers the Schwarzschild line element in harmonic form if the velocity is set to zero. In \cite{2014CoTPh..61..270H} the authors at first  write the Schwarzschild line element in terms of  harmonic coordinates.  Then they  write the Lorentz  transformation equations for  the inertial  frames  using  harmonic  co-ordinates,  which  is an approximation  (viz  their  Eqn. (7) ).  Actually the harmonic coordinate system used by those authors is the closest approximation available in general relativity to an inertial frame of reference in special relativity \cite{Dirac} (pp. 40-41).  So finally the authors in that paper used this approximated Lorentz transformation to obtain the line element of a uniformly moving Schwarzschild object.\\\\
In the contrary, in our present paper  we took an  alternate  approach.  We at first  write a far-field approximated Schwarzschild line element in Cartesian coordinates.   Then  we  consider  the  frame  in  which the  Schwarzschild object  is  placed,  is undergoing  uniform motion. Such form of the line element enables us to use the standard Lorentz transformation relations. So, in our case we used the standard Lorentz transformation in Cartesian form to obtain the line element of a  moving Schwarzschild object. Although strictly  speaking the specific properties of the parallel transport of vectors in Minkowski spacetime is not  allowed in Schwarzschild geometry  \cite{PhysRevD.99.064012}, but we  use the far-field approximated line element \eqref{x7},  which allow us to perform the Lorentz transformation in the present context as shown in \cite{ 1993ApJ...415..459P, PhysRevD.75.062002, PhysRevD.69.063001}.  \\\\
Since in our case we have used the standard Lorentz transformation relation in the Cartesian form of the far-field approximated Schwarzschild line element, we  don't  expect the result obtained in our paper to match exactly with the results obtained by the authors in \cite{2014CoTPh..61..270H} through harmonic coordinate transformation.

\section{Metric for a uniformly accelerated Schwarzschild object}
In the previous section we investigated the spacetime metric due to a uniformly moving Schwarzschild object. In this Section, we shall consider that the Schwarzschild object is in accelerated motion and analyze the geometry of the spacetime due to such an object. For that we describe the spacetime metric for an accelerated object in terms of an accelerated frame $\Bar{K} (c\Bar{t}, \Bar{x}, \Bar{y}, \Bar{z})$ which is moving with constant proper acceleration with respect to an inertial frame $K (ct, x, y, z)$. Meanwhile, the transformation relations between the frames $K (ct, x, y, z)$ and $\Bar{K} (c\Bar{t}, \Bar{x}, \Bar{y}, \Bar{z})$ where respective coordinate axes $X$ and $\Bar{X}$ of both the frames are coinciding and other coordinate axes $(Y, Z)$ and $(\Bar{Y}, \Bar{Z})$ are parallel to each other along with the constant proper acceleration $a_x$ in $X-$direction, are known as the Rindler transformation. Mathematically this transformation can be written as below \cite{David1997, Yi, wRindler2}:

\begin{equation}
    \begin{aligned}\label{eq:11}
    ct&=\left(\frac{c^2}{a_x}+\Bar{x}\right)\sinh \left(\frac{a_x\, \Bar{t}}{c}\right)\text{,\,\,\,\,\,\,\,}
    x=\left(\frac{c^2}{a_x}+\Bar{x}\right)\cosh \left(\frac{a_x\,\Bar{t}}{c}\right)-\frac{c^2}{a_x}\\
    y&=\Bar{y}
    \text{\,\,\,\,and\,\,\,\,}z=\Bar{z}
\end{aligned}
\end{equation}
The inverse transformation relation of equation \eqref{eq:11} can be written as:

\begin{equation}
    \begin{aligned}\label{eq:11a}
    c\Bar{t}&=\frac{c^2}{a_x}\tanh^{-1}\Bigg(\frac{ct}{x+\frac{c^2}{a_x}}\Bigg)\text{,\,\,\,\,\,\,\,}
    \Bar{x}=\sqrt{\left(x+\frac{c^2}{a_x}\right)^2-c^2t^2}-\frac{c^2}{a_x}\\
    \Bar{y}&=y
    \text{\,\,\,\,and\,\,\,\,}\Bar{z}=z
\end{aligned}
\end{equation}
Now, since in the present case the Schwarzschild object is in accelerated motion, we write equation \eqref{eq:x3} as (replacing $(ct', x', y', z')$ respectively by $(c\Bar{t}, \Bar{x}, \Bar{y}, \Bar{z})$ ) :
\begin{align}
    ds^2=\left(1-\frac{2m}{\Bar{r}-m}\right)c^2d\Bar{t}^2-\left(1+\frac{2m}{\Bar{r}-m}\right)\Big\{d\Bar{x}^2+d\Bar{y}^2+d\Bar{z}^2\Big\}\label{eq:16a}
\end{align}
where $\Bar{r}=\sqrt{\Bar{x}^2+\Bar{y}^2+\Bar{z}^2}$.\\\\
It is to be mentioned that the flatness of a spacetime line element can be analyzed by describing the metric in the form \cite{David1997}:
\begin{align}
    ds^2 = V^2(X)dT^2 -U^2(X)dX^2 - dY^2 -dZ^2\label{Eqx17}
\end{align}
where in order to make the  space to be Minkowskian  one must have  the derivative of the function $V$ with respect to $X$  to be proportional to $U$.  The  Rindler line  element as used  in the present  work  passes this  test. Thus, as stated before in the case of uniform motion, the Rindler transformation also cannot be used directly to the standard Schwarzschild line element. However, in our present case we assume a far-field approximation such that the spacetime geometry is close to flat, and hence we can use the said transformation relations on the approximated line element \eqref{eq:16a}. \\\\
Now using the transformation relation \eqref{eq:11a} we obtain from equation \eqref{eq:16a} as follows:
\begin{align}
    ds^2&=\Bigg[\frac{c^4}{a_x^2 \left\{\chi\right\}^2}\Big\{1-\frac{2 m}{\Delta_0}\Big\}\left( \frac{c^2}{a_x}+x\right)^2-\frac{c^2t^2 }{\chi}\left(1+\frac{2 m}{\Delta_0}\right)\Bigg]c^2dt^2\nonumber\\
    &-\Bigg[\frac{1}{\chi}\left(\frac{c^2}{a_x}+x\right)^2 \left(1+\frac{2 m}{\Delta_0}\right)-\frac{c^6 t^2 }{a_x^2\left\{\chi\right\}^2}\left(1-\frac{2 m}{\Delta_0}\right)\Bigg]dx^2\nonumber\\
    &-\Bigg[1+\frac{2 m}{\Delta_0}\Bigg]\Big\{dy^2+dz^2\Big\}+2\Bigg[\frac{ct }{\chi}\left(\frac{c^2}{a_x}+x\right) \left(1+\frac{2 m}{\Delta_0}\right)\nonumber\\
    &-\frac{c^6t }{a_x^2 \left\{\chi\right\}^2}\left(1-\frac{2 m}{\Delta_0}\right)\left(\frac{c^2}{a_x}+x\right)\Bigg]cdtdx\label{17}
\end{align}
where,
    \begin{align}
    \Delta_0&=\Bigg[\Bigg\{\chi^{1/2}-\frac{c^2}{a_x}\Bigg\}^2+y^2+z^2\Bigg]^{1/2}-m\,\,\,\,\text{and}\hspace{0.5cm}\chi=\left(\frac{c^2}{a_x}+x\right)^2 -c^2t^2
\end{align}
So equation \eqref{17}  represents the  line element for  an accelerated  Schwarzschild mass.
However, this  Schwarzschild  mass has  zero  initial velocity.  In order to accommodate  some  initial velocity in our dynamics  we  follow a  procedure in next sub-section.\\\\
Further, if we consider that the direction of the proper acceleration $a$ of an accelerated frame is in some arbitrary direction on the $XY$ plane, then it will have two components, one along the $X$ and other, along the $Y$ direction.  So, as the acceleration is in some arbitrary direction on the $XY$ plane then the inverse Rindler transformation relations in such a case can be written as (see {\bf Appendix A.2}) :
    \begin{equation}
    \begin{aligned}\label{k15}
    c\Bar{t}&=\frac{c^2}{a_{xy}}\tanh^{-1}\Bigg\{\frac{a_{xy}t}{c\Big(\frac{a_x}{c^2}x+\frac{a_y}{c^2}y+1\Big)}\Bigg\}\\
    \Bar{x}&=x+\Bigg[-\frac{1}{a_{xy}^2}\left(xa_x+ya_y\right)+\Bigg\{\left( \frac{1}{a_{xy}^2}\left(xa_x+ya_y\right)+\frac{c^2}{a_{xy}^2}\right)^2\\
    &\hspace{7cm}-\frac{c^2t^2}{a_{xy}^2}\Bigg\}^{1/2}-\frac{c^2}{a_{xy}^2}\Bigg]a_x\\
    \Bar{y}&=y+\Bigg[-\frac{1}{a_{xy}^2}\left(xa_x+ya_y\right)+\Bigg\{\left( \frac{1}{a_{xy}^2}\left(xa_x+ya_y\right)+\frac{c^2}{a_{xy}^2}\right)^2\\
    &\hspace{7cm}-\frac{c^2t^2}{a_{xy}^2}\Bigg\}^{1/2}-\frac{c^2}{a_{xy}^2}\Bigg]a_y\\
    &\text{and}\\
    \Bar{z}&=z
\end{aligned}
\end{equation}
where $a_x$ and $a_y$ respectively represent the $X$ and $Y$ components of the proper acceleration and $a_{xy}=\left(a_x^2+a_y^2\right)^{1/2}$. If the acceleration is only in $X-$direction then $a_y = 0$, $a_{xy} = a_x$ and then equation \eqref{k15} reduces to \eqref{eq:11a}.\\\\
Now in the above in both the transformation equations \eqref{eq:11a} and \eqref{k15}, the initial velocity of the accelerated frame is zero.

\subsection{Line element for a uniformly accelerated Schwarzschild object (with non-zero initial velocity) in $X-$direction}
The gravitational field due to an accelerated Schwarzschild object moving along $X$--direction can be described using the transformation equation \eqref{eq:11a}. However, the transformation relations between inertial (rest frame) and accelerated (non-inertial) frames with non-zero initial velocity can written as follows (see {\bf Appendix A.3}) \cite{Yi}:  
\begin{equation}
    \begin{aligned}\label{eq:x12}
    ct&=\gamma\Bigg[\left(\frac{c^2}{a_x}+\Bar{x}\right)\sinh \left(\frac{a_x\, \Bar{t}}{c}\right)+\frac{v_x}{c} \cdot \Bigg\{\left(\frac{c^2}{a_x}+\Bar{x}\right)\cosh \left(\frac{a_x\,\Bar{t}}{c}\right)-\frac{c^2}{a_x}\Bigg\}\Bigg] \\
    x&=\gamma\Bigg[\left(\frac{c^2}{a_x}+\Bar{x}\right)\cosh \left(\frac{a_x\,\Bar{t}}{c}\right)-\frac{c^2}{a_x}+\frac{v_x}{c}\,\left(\frac{c^2}{a_x}+\Bar{x}\right)\sinh \left(\frac{a_x\, \Bar{t}}{c}\right)\Bigg]\\
     y&=\Bar{y}\text{\,\,\,\,\,and\,\,\,\,\,\,}z=\Bar{z}
\end{aligned}
\end{equation}
where $\gamma=\left(1-\frac{v_x^2}{c^2}\right)^{-1/2}$ and $v_x$ is the initial velocity of the accelerated frame in $X-$direction.\\\\
At the boundary condition $a_x\longrightarrow 0$ the transformation relations \eqref{eq:x12} reduce to the standard Lorentz transformation and at $v_x=0$ the relations reduce to standard Rindler transformation relations \eqref{eq:11}.\\\\
Furthermore, the inverse transformation relation of \eqref{eq:x12} can be written as follows (see {\bf Appendix A.3}):
\begin{equation}
\begin{aligned}\label{xx24}
c\Bar{t}&=\frac{c^2}{a_x}\tanh^{-1}\left\{\frac{\left(ct+\gamma\frac{v_xc}{a_x}\right)-\frac{v_x}{c}\left(x+\gamma\frac{c^2}{a_x}\right)}{\left(x+\gamma\frac{c^2}{a_x}\right)-\frac{v_x}{c}\left(ct+\gamma\frac{v_xc}{a_x}\right) }\right\}\\
    \Bar{x}&=\Bigg[\left(x+\gamma\frac{c^2}{a_x}\right)^2-\left(ct+\gamma\frac{v_xc}{a_x}\right)^2\Bigg]^{1/2}-\frac{c^2}{a_x}\\
    \Bar{y}&=y\text{\,\,\,\,\,and\,\,\,\,} \Bar{z}=z
\end{aligned}
\end{equation}
Now in order to calculate the spacetime metric for an accelerated Schwarzschild object with
initial velocity $v_x$ and proper acceleration $a_x$ in $X-$direction we use the similar approach adopted in the previous section. So using equations in \eqref{xx24} in \eqref{eq:16a} and simplifying we obtain as below\footnote{We used { MATHEMATICA 13.2} during the calculation of equation \eqref{eq:x14}.}:
\begin{align}
    ds^2&=\Bigg[\frac{c^4 \left(a_x x+\gamma  c^2\right)^2 \left(\Delta_1-3 m\right)}{\left(\Delta_2\right)^2 \left(\Delta_1-m\right)}-\frac{\left(\frac{c \gamma  v_x}{a_x}+ct\right)^2 \left(1+\frac{2 m}{\Delta_1-m}\right)}{\left(\frac{\gamma  c^2}{a_x}+x\right)^2-\left(\frac{c \gamma  v_x}{a_x}+ct\right)^2}\Bigg]c^2dt^2\nonumber\\
    &-\Bigg[\frac{\left(\frac{\gamma  c^2}{a}+x\right)^2 \left(1+\frac{2 m}{\Delta_1-m}\right)}{\left(\frac{\gamma  c^2}{a_x}+x\right)^2-\left(\frac{c \gamma  v_x}{a_x}+ct\right)^2}-\frac{c^4 (a_x ct+c \gamma  v_x)^2 \left(\Delta_1-3 m\right)}{\left(\Delta_2\right)^2 \left(\Delta_1-m\right)}\Bigg]dx^2\nonumber\\
    &-\Bigg[1+\frac{2 m}{\Delta_1-m}\Bigg]\Big\{dy^2+dz^2\Big\}+2\Bigg[\frac{\left(\frac{\gamma  c^2}{a_x}+x\right) \left(\frac{c \gamma  v_x}{a_x}+ct\right) \left(1+\frac{2 m}{\Delta_1-m}\right)}{\left(\frac{\gamma  c^2}{a_x}+x\right)^2-\left(\frac{c \gamma  v_x}{a_x}+ct\right)^2}\nonumber\\
    &-\frac{c^4 \left(a_x x+\gamma  c^2\right) (a_x ct+c \gamma  v_x) \left(\Delta_1-3 m\right)}{\left(\Delta_2\right)^2 \left(\Delta_1-m\right)}\Bigg]cdtdx\label{eq:x14}
\end{align}
where,
\begin{align}
    \Delta_1&=\Bigg\{\Bigg\{\Bigg\{\left(\frac{\gamma  c^2}{a_x}+x\right)^2-\left(\frac{c \gamma  v_x}{a_x}+ct\right)^2\Bigg\}^{1/2}-\frac{c^2}{a_x}\Bigg\}^2+y^2+z^2\Bigg\}^{1/2}\\
    \Delta_2&=a_x^2 \left(c^2t^2-x^2\right)+2 a_x c \gamma  (ct v_x-c x)+c^2 \gamma ^2 \left(v_x^2-c^2\right)
\end{align}
The equation \eqref{eq:x14} is the spacetime line element for an accelerated Schwarzschild object with initial velocity $v_x$ and proper acceleration $a_x$ in $X-$direction. \\\\
Below we analyze the line element \eqref{eq:x14} at different limiting conditions.
\begin{enumerate}
    \item[(a)] \textbf{Case-I:} If $a_x\longrightarrow 0$  we obtain from equation \eqref{eq:x14} as follows:
\begin{align}
 ds^2&= \Bigg[1-\frac{2 m\gamma ^2}{\mathcal{R}}\Big\{1+\frac{v_x^2}{c^2}\Big\}\Bigg]c^2dt^2- \Bigg[1+\frac{2 m\gamma ^2}{\mathcal{R}}\Big\{1+\frac{v_x^2}{c^2}\Big\}\Bigg]dx^2\nonumber\\
    &-\Big[1+\frac{2 m}{\mathcal{R}}\Big]\Big\{dy^2+dz^2\Big\}+\frac{8 \gamma ^2 m v_x}{\mathcal{R}}\,dx dt \label{eq:16}
\end{align}
At $a_x\longrightarrow 0$ the equation \eqref{eq:x14} reduces to \eqref{eq:x10}  that represents the line element due to a uniformly moving Schwarzschild object.\\
\item[(b)] \textbf{Case-II:} If $v_x\longrightarrow0$ then equation \eqref{eq:x14} reduces to the line element as below:
\begin{align}
  ds^2&=\Bigg[\frac{c^4}{a_x^2 \left\{\chi\right\}^2}\Big\{1-\frac{2 m}{\Delta_0}\Big\}\left( \frac{c^2}{a_x}+x\right)^2-\frac{c^2t^2 }{\chi}\left(1+\frac{2 m}{\Delta_0}\right)\Bigg]c^2dt^2\nonumber\\
    &-\Bigg[\frac{1}{\chi}\left(\frac{c^2}{a_x}+x\right)^2 \left(1+\frac{2 m}{\Delta_0}\right)-\frac{c^6 t^2 }{a_x^2\left\{\chi\right\}^2}\left(1-\frac{2 m}{\Delta_0}\right)\Bigg]dx^2\nonumber\\
    &-\Bigg[1+\frac{2 m}{\Delta_0}\Bigg]\Big\{dy^2+dz^2\Big\}+2\Bigg[\frac{ct }{\chi}\left(\frac{c^2}{a_x}+x\right) \left(1+\frac{2 m}{\Delta_0}\right)\nonumber\\
    &-\frac{c^6t }{a_x^2 \left\{\chi\right\}^2}\left(1-\frac{2 m}{\Delta_0}\right)\left(\frac{c^2}{a_x}+x\right)\Bigg]cdtdx
\label{eq:15}
\end{align} 
The line element \eqref{eq:15} is exactly same as in \eqref{17} that represents the spacetime geometry due to an accelerated Schwarzschild object (accelerated along the $X-$ direction) with zero initial velocity.\\
\item[(c)] \textbf{Case-III:} At the limits $v_x\longrightarrow 0$ and $a_x\longrightarrow 0$ we obtain from equation \eqref{eq:x14} as below:
\begin{align}
    ds^2&=\left(1-\frac{2m}{r-m}\right)c^2dt^2-\left(1+\frac{2m}{r-m}\right)\Big\{dx^2+dy^2+dz^2\Big\}\label{eq:17}
\end{align}
which is exactly the Schwarzschild metric in Cartesian coordinates shown in \eqref{eq:x3} (since $ct = ct'$, $ x = x'$, $y=y'$, $z=z'$ and $r=r'$ for $v_x\longrightarrow 0$ and $a_x\longrightarrow 0$).\\
\item[(d)] \textbf{Case-IV:} If $v_x\longrightarrow 0$ and $m\longrightarrow 0$ then equation \eqref{eq:x14} reduces to
\begin{align}
   ds^2&=\Bigg[\frac{c^4}{a_x^2 \left\{\chi\right\}^2}\left( \frac{c^2}{a_x}+x\right)^2-\frac{c^2t^2 }{\chi}\Bigg]c^2dt^2-\Bigg[\frac{1}{\chi}\left(\frac{c^2}{a_x}+x\right)^2 -\frac{c^6 t^2 }{a_x^2\left\{\chi\right\}^2}\Bigg]dx^2\nonumber\\
    &-\Big\{dy^2+dz^2\Big\}+2\Bigg[\frac{ct }{\chi} -\frac{c^6t }{a_x^2 \left\{\chi\right\}^2}\Bigg]\left(\frac{c^2}{a_x}+x\right)cdtdx\label{Eqx27}
\end{align}
Equation \eqref{Eqx27} is a metric in an accelerated frame as observed from a rest frame. 

\end{enumerate}
From the above analysis, we can confirm that as the obtained line element \eqref{eq:x14} fulfills all the necessary limiting conditions, so this line element can represent the gravitational field due to an accelerated Schwarzschild object (with some  initial velocity in the same  direction).

\subsection{Line element for a Schwarzschild object having constant acceleration in some arbitrary direction on the $XY$ plane}
The transformation relation for an accelerated frame with respect to an inertial frame $K$ where the acceleration is in arbitrary direction on the $XY$ plane having zero initial velocity is shown in equation \eqref{k15} (also see {\bf Appendix A.2}). So using this transformation relation in equation \eqref{eq:16a} we obtain the line element for a Schwarzschild object having acceleration in some arbitrary direction on the $XY$ plane as below:
\begin{align}
    ds^2&=g_{t\,t}c^2dt^2+g_{x\,x}dx^2+g_{y\,y}dy^2+g_{z\,z}dz^2\nonumber\\
    &\hspace{1cm}+2\Big\{g_{x\,t}dxcdt+g_{y\,t}dycdt+g_{x\,y}dxdy\Big\}\label{k26}
\end{align}
where the metric components can be written as follows \footnote{We used { MATHEMATICA 13.2} during the calculation of metric elements \eqref{q27}--\eqref{q36}.}:

\begin{align}
    g_{t\,t}&=\frac{c^4 }{ \left(\xi_2\right)^2}\left(1-\frac{2 m}{\Delta_3-m}\right)\left(a_x x+a_y y+c^2\right)^2\nonumber\\
    &\hspace{2cm}- \frac{ c^2t^2}{\xi_2}\left(1+\frac{2 m}{\Delta_3-m}\right)\Big\{a_x^2 +a_y^2 \Big\}\label{q27}\\
    g_{x x}&=-\left(1+\frac{2 m}{\Delta_3-m}\right)\Bigg[\left(\frac{a_x^2 \xi_1}{a^2}+1\right)^2+\frac{a_x^2 a_y^2 \xi_1^2 }{a^4}\Bigg]\nonumber\\
    &\hspace{4cm}+\frac{a_x^2 c^6 t^2}{\left(\xi_2\right)^2} \left(1-\frac{2 m}{\Delta_3-m}\right)\\
    g_{yy}&=-\left(1+\frac{2 m}{\Delta_3-m}\right)\Bigg[\left(\frac{a_y^2 \xi_1}{a^2}+1\right)^2 +\frac{a_x^2 a_y^2 \xi_1^2 }{a^4}\Bigg]\nonumber\\
    &\hspace{2cm}+\frac{a_y^2 c^6 t^2 }{\left(\xi_2\right)^2}\left(1-\frac{2 m}{\Delta_3-m}\right)\\
    g_{zz}&=-\Bigg\{1+\frac{2 m}{\Delta_3-m}\Bigg\}\\
    g_{xt}&=a_x ct \Bigg[\frac{1}{\sqrt{\xi_2}}\Bigg\{\frac{ \xi_1}{a^2} \Big(a_x^2+a_y^2 \Big)+1\Bigg\}\left(1+\frac{2 m}{\Delta_3-m}\right)\nonumber\\
    &-\frac{c^5 }{\left(\xi_2\right)^2}\left(1-\frac{2 m}{\Delta_3-m}\right)\left(a_x x+a_y y+c^2\right)\Bigg]\\
    g_{yt}&=a_y ct \Bigg[\frac{1}{ \sqrt{\xi_2}}\Bigg\{\frac{ \xi_1}{a^2}\Big(a_x^2 +a_y^2 \Big)+1\Bigg\}\left(1+\frac{2 m}{\Delta_3-m}\right)\nonumber\\
    &-\frac{c^5}{\left(\xi_2\right)^2} \left(1-\frac{2 m}{\Delta_3-m}\right)\left(a_x x+a_y y+c^2\right)\Bigg]\\
    g_{xy}&=a_x a_y \Bigg[\frac{c^6 t^2 }{ \left(\xi_2\right)^2}\left(1-\frac{2 m}{\Delta_3-m}\right)-\left(1+\frac{2 m}{\Delta_3-m}\right)\nonumber\\
    &\hspace{2cm}\times\frac{\xi_1 }{a^2}\Bigg\{\frac{ \xi_1}{a^2} \left(a_x^2+ a_y^2 \right)+2 \Bigg\}\Bigg]\label{q36}
\end{align}
in which $\Delta_3$, $\xi_1$ and $\xi_2$ are used to represent as follows:
\begin{align}
    \Delta_3&=\Bigg\{\Big\{ x+ \frac{a_x}{a^2}\Big( \sqrt{\xi_2}-a_x x- a_y y- c^2\Big)\Big\}^2\nonumber\\
    &+\Big\{ y+ \frac{a_y}{a^2} \Big(\sqrt{\xi_2}-a_x x-a_y y- c^2\Big)\Big\}^2+ z^2\Bigg\}^{1/2}\\
    \xi_1&=\left\{\frac{1}{ \sqrt{\xi_2}}\Big(a_x x+a_y y+c^2\Big)-1\right\}\\
    \text{and\,\,\,\,}\xi_2&=\left(a_x x+a_y y+c^2\right)^2-a^2 c^2t^2
\end{align}
The above metric \eqref{k26} (with metric components \eqref{q27}-\eqref{q36}) recovers the metric \eqref{eq:x14} for a Schwarzschild object having acceleration in $X-$direction, if the $Y$ component of the acceleration is set to zero, i.e. $a_y\longrightarrow 0$.

\section{Discussion of results and conclusions}
In this work, we investigated the gravitational field due to a moving Schwarzschild object, where we considered uniform accelerated motion in $X-$direction. And  we  also established  the transformation  equations,  after   decomposing the  given acceleration  into two  arbitrary  orthogonal directions ( which   may be  useful  for  some future work). Subsequently, we obtained the spacetime line element for each cases. Our findings can be described as follows:
\begin{enumerate}
    \item[(a)] At first we obtained the line element for the spacetime geometry due to a uniformly moving Schwarzschild object. If the velocity is  set to zero as a limiting condition, then we recover the standard Schwarzschild line element. Further, this line element becomes Minkowskian line element if the mass parameter is set to zero which indicates the flatness of the spacetime. 
    \item[(b)] Next, we extended our work for an object having uniformly accelerated motion-- (i) acceleration in $X-$direction and (ii) acceleration in an arbitrary direction on the $XY$ plane. In both the cases we obtained the spacetime line elements and we  successfully tested those line elements under various limiting conditions.
       
\end{enumerate}
The spacetime geometries (in terms of the line elements) described above are well defined and satisfy their respective necessary limiting conditions. Therefore we concluded that the obtained line elements can represent the respective gravitational fields due to the moving Schwarzschild objects with various kinds of motions.\\\\
Such work will have relevant applications in astrophysics when we calculate the geodesic equations of test particles in the gravitational field of objects having uniform acceleration. For  example, our  calculations  will be useful  in a situation where we have  electrons and  other particles  moving in the Roche lobes  of binaries. Here each component  of the  binary  is  having its  own gravitational  field  and  at  the same time is in an accelerated  state  of  motion.   Also  in the case of collision/ merging of two black  holes,   we  have  gravitational  field  produced  by  objects  which  themselves  are  in accelerated  state  of  motion. When two black holes merge, the remnant black hole can receive a kick or recoil. This is because of the asymmetric emission of gravitational waves during the merger process. Since gravitational waves carry away linear momentum, the remnant black hole recoils in the opposite direction. The kick is primarily caused by the anisotropic emission of gravitational waves, especially when the black holes have unequal masses or misaligned spins. The recoil velocity can reach up to 5000 km/sec. There is possible evidence that the merger event GW200129 provided strong evidence of a large kick velocity, estimated to be around 1500 km/s \cite{PhysRevLett.128.191102}.
Besides, if  we want to calculate  the trajectory of  a light ray (  or  dynamics  of  a test particle)   around  such an  accelerated  gravitational body,  the set  of  calculations  done  in our present work  would be useful. Besides  our calculations  will be useful  in the framework  of  cosmology,  where  dynamics takes  place  under the influence  of  dark  matter and  dark energy. Further  our calculations  can be  extended  to find  the  gravitational  field  due  to  rotating objects.  And such  calculations  will be  useful  in the studies  of  microlensing and  detecting exoplanets. \\\\
Most  importantly,  our tools  will  be  useful   in the studies of gravitational waves. The gravitational-wave signal observed by a detector in the vicinity of Earth   is  actually approximated by the waves along some geodesic related to distance from the source.   Thus, if we   estimate  a quantity  as  detected by the Gravitational Wave (GW)  detector, we approximate the signal an inertial observer measures as a function of proper time \cite{PhysRevD.93.084031}.  Furthermore, the Rindler framework can be used to study the behavior of detectors in curved spacetimes, like those around black holes, where the spacetime curvature is significant.    In the context of GW detection, the Rindler transformation can be used to analyze the interaction between GWs and detectors, particularly when considering the quantum nature of both.   The Rindler transformation provides a valuable tool for understanding the perspectives of accelerated observers and the thermal effects associated with horizons in both flat and curved spacetime. This, in turn, helps in exploring complex phenomena like the behavior of GW detectors in the presence of strong gravitational fields and the interplay between gravitational waves, quantum fields, and the Unruh effect.  The Rindler transformation helps us understand how GWs can transfer energy to the internal states of a detector, potentially affecting its coherence and response  \cite{Barros2024}.  
By studying the response of detectors in the Rindler frame, researchers can gain insights into the fundamental nature of quantum fields in curved spacetime and the properties of GWs themselves \cite{chowdhury2024finitetemperaturecftrindlervacuum}. 
The Rindler transformation  in the background  of curved  spacetime,  provides a valuable theoretical tool for bridging the gap between general relativity, quantum field theory, and the experimental realm of GW detection \cite{Liu2023}.  Thus  our present  study  on how  Lorentz frame /  Rindler   frame  behaves  in the  background  of  curved  spacetime (in our case  Schwarzschild field)   will have relevance in the research  on  gravitational wave  and its  detection .

\section*{Appendix}
\appendix
\numberwithin{equation}{section}
\section{Transformation relations for uniformly accelerated frames}

\subsection{Inverse Rindler transformation}
The standard Rindler transformation relation for acceleration along the $X-$direction can be written as follows \cite{David1997}:
    \begin{align}
    ct&=\left(\frac{c^2}{a_x}+\Bar{x}\right)\sinh \left(\frac{a_x\, \Bar{t}}{c}\right),\hspace{0.5cm}
    x=\left(\frac{c^2}{a_x}+\Bar{x}\right)\cosh \left(\frac{a_x\,\Bar{t}}{c}\right)-\frac{c^2}{a_x}\label{e2}
\end{align}
Further, from the above equations, we can obtain as:
\begin{align}
&ct=\sinh\left(\frac{a_x\,\Bar{t}}{c}\right)\frac{c^2}{a_x}\Bigg\{1+\frac{a_x}{c^2}\,\Bar{x}\Bigg\}\label{e3}\\
  \text{and\,\,\,\,}  &\frac{c^2}{a_x}\left(1+\frac{a_x}{c^2}\,x\right)=\frac{c^2}{a_x}\cosh\left(\frac{a_x\,\Bar{t}}{c}\right)\Bigg\{1+\frac{a_x}{c^2}\,\Bar{x}\Bigg\}\label{e4}
\end{align}
Now, dividing equation \eqref{e3} by \eqref{e4} we obtain:
\begin{align}
   & \frac{ct}{\frac{c^2}{a_x}\left(1+\frac{a_x}{c^2}\,x\right)}=\tanh\left(\frac{a_x\,t}{c}\right)
   \text{\,\,\,\,\,\,or }\,\,\,\,c\Bar{t}=\frac{c^2}{a_x}\tanh^{-1}\left[\frac{ct}{\frac{c^2}{a_x}+x}\right]\label{e6}
\end{align}
Again, from equations \eqref{e3} and \eqref{e4} we can write as:
\begin{align}
    \Bigg[\frac{c^2}{a_x}\left(1+\frac{a_x}{c^2}\,x\right)\Bigg]^2-c^2t^2&=\left(\frac{c^2}{a_x}\right)^2\Bigg\{1+\frac{a_x}{c^2}\,\Bar{x}\Bigg\}^2\nonumber\\&\times\Bigg[\cosh^2\left(\frac{a_x\,\Bar{t}}{c}\right)-\sinh^2\left(\frac{a_x\,\Bar{t}}{c}\right)\Bigg]\nonumber\\
    \text{or\,\,\,\,\,\,}\left(\frac{c^2}{a_x}\right)\Bigg\{1+\frac{a_x}{c^2}\,\Bar{x}\Bigg\}&=\sqrt{\Bigg[\frac{c^2}{a_x}\left(1+\frac{a_x}{c^2}\,x\right)\Bigg]^2-c^2t^2}\nonumber\\
    \text{or\,\,\,\,\,\,}\Bar{x}&=\sqrt{\Bigg[\left(\frac{c^2}{a_x}+x\right)\Bigg]^2-c^2t^2}-\frac{c^2}{a_x}\label{e8}
\end{align}
Equations \eqref{e6} and \eqref{e8} represents the inverse Rindler transformation between an accelerated frame $\Bar{K}$ and a rest frame $K$ along the $X-$direction.

\subsection{Inverse Rindler transformation for acceleration in arbitrary direction}

Here we follow a procedure similar to that of the previous work \cite{Sangwha2020}. Consider that $\vec{r}$ is the position vector of a point $P$ in a reference frame $K(ct, x, y, z)$ which can be written as a sum of a parallel vector $\Vec{r}_{\parallel}$ and a perpendicular vector $\Vec{r}_{\perp}$ as follows (see {\bf Fig. 1}) :
\begin{align}
    \Vec{r}=\Vec{r}_{\parallel}+\Vec{r}_{\perp}
\end{align}
\begin{figure}[hpt]
\centering
\minipage{0.5\textwidth}
\includegraphics[width=\linewidth]{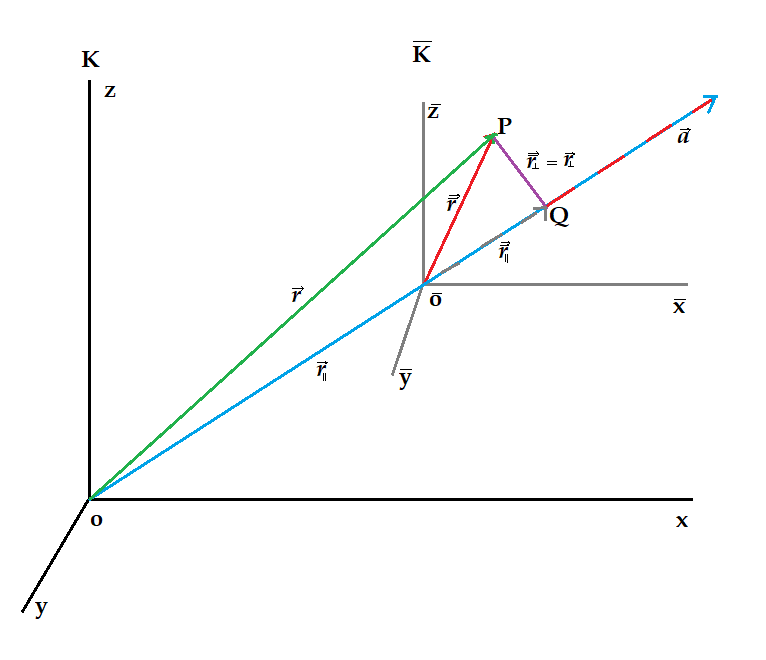}
\caption{\textit{Accelerated frame $\Bar{K}$ moving with respect to a rest frame $K$ in some arbitrary direction.}}
\label{Fig:Qm1}
\endminipage
\end{figure}
Further the parallel and the perpendicular vectors can be written as:
\begin{align}
    \Vec{r}_{\parallel}&=\frac{\left(\Vec{r}\cdot\Vec{a}\right)}{\vert\Vec{a}\vert^2}\Vec{a}\,\,\hspace{1cm}
    \text{and\,\,\,\,\,\,\,}\Vec{r}_{\perp}=\Vec{r}-\frac{\left(\Vec{r}\cdot\Vec{a}\right)}{\vert\Vec{a}\vert^2}\Vec{a}
\end{align}
where $\vec{a}$ is the acceleration in the direction of  $\Vec{r}_{\parallel}$. Now we consider another frame $\Bar{K}(c\Bar{t}, \Bar{x}, \Bar{y}, \Bar{z})$ where the position vector of the point $P$ in that frame is $\Vec{\Bar{r}}$ which also can be written as:
\begin{align}
    \Vec{\Bar{r}}=\Vec{\Bar{r}}_{\parallel}+\Vec{\Bar{r}}_{\perp}
\end{align}
where the $\Vec{\Bar{r}}_{\parallel}$ is along the direction of $\Vec{a}$  and $\Vec{\Bar{r}}_{\perp}$ is perpendicular to $\Vec{a}$.\\\\
Then the transformation relation between the vectors $\Vec{\Bar{r}}_{\parallel}$ and $\Vec{r}_{\parallel}$ can be written by using the standard inverse Rindler transformation as below (as the acceleration $\Vec{a}$ of the frame $\Bar{K}$ is in the direction of $\Vec{r}_{\parallel}$):{\bf
\begin{align}
 \Vec{\Bar{r}}_{\parallel}&=\Bigg[\sqrt{\left( \frac{\left(\Vec{r}\cdot\Vec{a}\right)}{\vert\Vec{a}\vert^2}+\frac{c^2}{\vert\vec{a}\vert^2}\right)^2-\frac{c^2t^2}{\vert\vec{a}\vert^2}}-\frac{c^2}{\vert\vec{a}\vert^2}\Bigg]\vec{a}\label{B24}
\end{align}
}\\
And since the vectors $\Vec{\Bar{r}}_{\perp}$ and $\Vec{r}_{\perp}$ are perpendicular to the acceleration $\vec{a}$, so the acceleration has no effect on those vectors and hence
\begin{align}\label{B25}
    \Vec{\Bar{r}}_{\perp}=\Vec{r}_{\perp}
\end{align}
Further as the acceleration of the frame $\Bar{K}$ is in the direction of the vector $\Vec{r}_{\parallel}$ the inverse time coordinate transformation between the frames $\Bar{K}$ and $K$ can be written as:{\bf
\begin{align}
    c\Bar{t}&=\frac{c^2}{a}\tanh^{-1}\Bigg\{\frac{a\,t}{c\Big(\frac{\vec{a}}{c^2}\cdot \Vec{r}_{\parallel}+1\Big)}\Bigg\}=\frac{c^2}{a}\tanh^{-1}\Bigg\{\frac{a\,t}{c\Big\{\frac{\vec{a}}{c^2}\cdot \left(\Vec{r}_{\parallel}+\Vec{r}_{\perp}\right)+1\Big\}}\Bigg\}\nonumber\\
    &=\frac{c^2}{a}\tanh^{-1}\Bigg\{\frac{a\,t}{c\Big(\frac{\vec{a}}{c^2}\cdot \Vec{r}+1\Big)}\Bigg\}\label{eq:xy36}
\end{align}
}\\
and the spatial coordinate transformation can be obtained from \eqref{B24} and \eqref{B25} (by putting $\vert\Vec{a}\vert=a$):{\bf
\begin{align}
    \Vec{\Bar{r}}&=\Vec{\Bar{r}}_{\perp}+\Vec{\Bar{r}}_{\parallel}=\Vec{r}_{\perp}+\Bigg[\sqrt{\left( \frac{\left(\Vec{r}\cdot\Vec{a}\right)}{\vert\Vec{a}\vert^2}+\frac{c^2}{\vert\vec{a}\vert^2}\right)^2-\frac{c^2t^2}{\vert\vec{a}\vert^2}}-\frac{c^2}{\vert\vec{a}\vert^2}\Bigg]\vec{a}\nonumber\\
    &=\Vec{r}-\frac{\left(\Vec{r}\cdot\Vec{a}\right)}{\vert\Vec{a}\vert^2}\Vec{a}+\Bigg[\sqrt{\left( \frac{\left(\Vec{r}\cdot\Vec{a}\right)}{\vert\Vec{a}\vert^2}+\frac{c^2}{\vert\vec{a}\vert^2}\right)^2-\frac{c^2t^2}{\vert\vec{a}\vert^2}}-\frac{c^2}{\vert\vec{a}\vert^2}\Bigg]\vec{a}\nonumber\\
    &=\Vec{r}-\Bigg[\frac{\left(\Vec{r}\cdot\Vec{a}\right)}{a}-\Bigg\{\left( \frac{\left(\Vec{r}\cdot\Vec{a}\right)}{a}+\frac{c^2}{a}\right)^2-c^2t^2\Bigg\}^{1/2}+\frac{c^2}{a}\Bigg]\frac{\vec{a}}{a}\label{eq:xy37}
\end{align}
}\\
The equations \eqref{eq:xy36} and \eqref{eq:xy37} are the Rindler transformation of coordinates in arbitrary direction. \\\\
Now we consider that $\vec{a_x}$, $\Vec{a_y}$ and $\Vec{a_z}$ respectively are the $X$, $Y$ and $Z$ components of the acceleration $\Vec{a}$ such that $a=\sqrt{a_x^2+a_y^2+a_z^2}$. Then we obtain from eqn. \eqref{eq:xy36} as below:
\begin{align}
    c\Bar{t}&=\frac{c^2}{a}\tanh^{-1}\Bigg\{\frac{a\,t}{c\Big(\frac{a_x}{c^2}\,x+\frac{a_y}{c^2}\,y+\frac{a_z}{c^2}\,z+1\Big)}\Bigg\}
\end{align}
and from \eqref{eq:xy37}:
\begin{align}
   \vec{\Bar{x}}+&\vec{\Bar{y}}+\vec{\Bar{z}} =\Vec{x}+\Vec{y}+\Vec{z}-\Bigg[\frac{\left(\Vec{r}\cdot\Vec{a}\right)}{a}-\Bigg\{\left( \frac{\left(\Vec{r}\cdot\Vec{a}\right)}{a}+\frac{c^2}{a}\right)^2-c^2t^2\Bigg\}^{1/2}+\frac{c^2}{a}\Bigg]\frac{\vec{a}}{a}\nonumber\\
   &=\Vec{x}+\Bigg[-\frac{1}{a^2}\left(x\,a_x+y\,a_y+z\,a_z\right)+\Bigg\{\left( \frac{1}{a^2}\left(x\,a_x+y\,a_y+z\,a_z\right)+\frac{c^2}{a^2}\right)^2\nonumber\\
   &-\frac{c^2t^2}{a^2}\Bigg\}^{1/2}-\frac{c^2}{a^2}\Bigg]\Vec{a}_x+\Vec{y}+\Bigg[-\frac{1}{a^2}\left(x\,a_x+y\,a_y+z\,a_z\right)\nonumber\\
   &+\Bigg\{\left( \frac{1}{a^2}\left(x\,a_x+y\,a_y+z\,a_z\right)+\frac{c^2}{a^2}\right)^2-\frac{c^2t^2}{a^2}\Bigg\}^{1/2}-\frac{c^2}{a^2}\Bigg]\Vec{a}_y\nonumber\\
   &+\Vec{z}+\Bigg[-\frac{1}{a^2}\left(x\,a_x+y\,a_y+z\,a_z\right)+\Bigg\{\left( \frac{1}{a^2}\left(x\,a_x+y\,a_y+z\,a_z\right)+\frac{c^2}{a^2}\right)^2\nonumber\\
   &-\frac{c^2t^2}{a^2}\Bigg\}^{1/2}-\frac{c^2}{a^2}\Bigg]\Vec{a}_z
\end{align}
Now equating the $X$, $Y$ and $Z$ components and dropping the vector notation we obtain as:
    \begin{align}
    \Bar{x}&=x+\Bigg[-\frac{1}{a^2}\left(x\,a_x+y\,a_y+z\,a_z\right)\nonumber\\
    &+\Bigg\{\left( \frac{1}{a^2}\left(x\,a_x+y\,a_y+z\,a_z\right)+\frac{c^2}{a^2}\right)^2-\frac{c^2t^2}{a^2}\Bigg\}^{1/2}-\frac{c^2}{a^2}\Bigg]a_x\\
    \Bar{y}&=y+\Bigg[-\frac{1}{a^2}\left(x\,a_x+y\,a_y+z\,a_z\right)\nonumber\\
    &+\Bigg\{\left( \frac{1}{a^2}\left(x\,a_x+y\,a_y+z\,a_z\right)+\frac{c^2}{a^2}\right)^2-\frac{c^2t^2}{a^2}\Bigg\}^{1/2}-\frac{c^2}{a^2}\Bigg]a_y\\
    \text{and\,\,\,\,}\Bar{z}&=z+\Bigg[-\frac{1}{a^2}\left(x\,a_x+y\,a_y+z\,a_z\right)\nonumber\\
    &+\Bigg\{\left( \frac{1}{a^2}\left(x\,a_x+y\,a_y+z\,a_z\right)+\frac{c^2}{a^2}\right)^2-\frac{c^2t^2}{a^2}\Bigg\}^{1/2}-\frac{c^2}{a^2}\Bigg]a_z\label{eq:xy41}
\end{align}
where $a=\sqrt{a_x^2+a_y^2+a_z^2}$.\\\\
Now if the acceleration is on the $XY$ plane i.e. $a_z=0$ and $a=\sqrt{a_x^2+a_y^2}=a_{xy}$ then we can write the above transformation as:
\begin{align}
    c\Bar{t}&=\frac{c^2}{a_{xy}}\tanh^{-1}\Bigg\{\frac{a_{xy}\,t}{c\Big(\frac{a_x}{c^2}\,x+\frac{a_y}{c^2}\,y+1\Big)}\Bigg\}\\
    \Bar{x}&=x+\Bigg[-\frac{1}{a_{xy}^2}\left(x\,a_x+y\,a_y\right)+\Bigg\{\left( \frac{1}{a_{xy}^2}\left(x\,a_x+y\,a_y\right)+\frac{c^2}{a_{xy}^2}\right)^2\nonumber\\
    &\hspace{6cm}-\frac{c^2t^2}{a_{xy}^2}\Bigg\}^{1/2}-\frac{c^2}{a_{xy}^2}\Bigg]a_x\\
    \Bar{y}&=y+\Bigg[-\frac{1}{a_{xy}^2}\left(x\,a_x+y\,a_y\right)+\Bigg\{\left( \frac{1}{a_{xy}^2}\left(x\,a_x+y\,a_y\right)+\frac{c^2}{a_{xy}^2}\right)^2\nonumber\\
    &\hspace{6cm}-\frac{c^2t^2}{a_{xy}^2}\Bigg\}^{1/2}-\frac{c^2}{a_{xy}^2}\Bigg]a_y\\
    &\text{and\,\,\,\,}\nonumber\\
    \Bar{z}&=z
\end{align}
If the acceleration is only along the $X-$direction then $a_y=a_z=0$, $a=a_x$, then the above transformation relations become:
\begin{align}
c\Bar{t}&=\frac{c^2}{a_x}\tanh^{-1}\Bigg\{\frac{c\,t}{\Big(\frac{c^2}{a_x}+x\Big)}\Bigg\}\\
     \Bar{x}&=\Bigg\{\left( x+\frac{c^2}{a_x}\right)^2-c^2t^2\Bigg\}^{1/2}-\frac{c^2}{a_x}\\
    \Bar{y}&=y\\
    \text{and\,\,\,\,}\Bar{z}&=z
\end{align}
which is the standard inverse Rindler transformation relation. 

\subsection{Rindler transformation for the frames with non-zero initial velocity}
Consider a frame $K'(ct', x', y', z')$ is moving with uniform velocity $v_x$ along the $X-$direction with respect to a rest frame $K (ct, x, y, z)$, then the coordinate transformation between those frames can be written as:
\begin{equation}
    \begin{aligned}\label{be29}
    ct&=\gamma\left(ct'+\frac{v_x}{c} \cdot x'\right),\hspace{0.5cm}
    x=\gamma\left(x'+v_x\,t'\right), \hspace{0.5cm}y=y'\,\,,\hspace{0.5cm}
     z=z'
\end{aligned}
\end{equation}
where $\gamma=\left(1-v_x^2/c^2\right)^{-1/2}$ is the well known Lorentz factor.\\\\
Now, consider another frame $\Bar{K}(c\Bar{t}, \Bar{x}, \Bar{y}, \Bar{z})$ that is moving with uniform acceleration $(a_x)$ along the $X-$direction with respect to an inertial frame $K'(ct', x', y', z')$ where the frame $\Bar{K}$ has zero initial velocity. Then, the coordinate transformation relation between the frames $K'$ and $\Bar{K}$ can be represented as:
\begin{equation}
    \begin{aligned}\label{be30}
    ct'&=\left(\frac{c^2}{a_x}+\Bar{x}\right)\sinh \left(\frac{a_x\, \Bar{t}}{c}\right)\text{,\,\,\,\,}
    x'=\left(\frac{c^2}{a_x}+\Bar{x}\right)\cosh \left(\frac{a_x\,\Bar{t}}{c}\right)-\frac{c^2}{a_x}\\
    y'&=\Bar{y}
    \text{\,\,\,\,and\,\,\,\,}z'=\Bar{z}
\end{aligned}
\end{equation}
Here it is to be noted that as stated above the accelerated frame $\Bar{K}$ has zero initial velocity with respect to the frame $K'$, however the frame $K'$ itself is moving with uniform velocity $v_x$, therefore according to the observer in the rest frame $K$, the accelerated frame has non-zero initial velocity. The coordinate transformation relation between the inertial rest frame $K$ and the accelerated frame $\Bar{K}$ can be obtained by substituting the expressions of the coordinates in equation \eqref{be30} into equation \eqref{be29} as follows:
\begin{small}
    \begin{align}
    ct&=\gamma\Bigg[\left(\frac{c^2}{a_x}+\Bar{x}\right)\sinh \left(\frac{a_x\, \Bar{t}}{c}\right)+\frac{v_x}{c} \cdot \Bigg\{\left(\frac{c^2}{a_x}+\Bar{x}\right)\cosh \left(\frac{a_x\,\Bar{t}}{c}\right)-\frac{c^2}{a_x}\Bigg\}\Bigg] \label{be31}\\
    x&=\gamma\Bigg[\left(\frac{c^2}{a_x}+\Bar{x}\right)\cosh \left(\frac{a_x\,\Bar{t}}{c}\right)-\frac{c^2}{a_x}+\frac{v_x}{c}\,\left(\frac{c^2}{a_x}+\Bar{x}\right)\sinh \left(\frac{a_x\, \Bar{t}}{c}\right)\Bigg]\label{be32}\\
     y&=\Bar{y}\\
     z&=\Bar{z}
\end{align}
\end{small}
The above equation in the transformation relation of an accelerated frame having non-zero initial velocity. The inverse of this equation can be obtained as follows:\\\\
Equations \eqref{be31} and \eqref{be32} can be rewritten as:
\begin{align}
    \left(ct+\gamma\frac{v_xc}{a_x}\right)&=\gamma\left(\frac{c^2}{a_x}+\Bar{x}\right)\Bigg[\sinh \left(\frac{a_x\, \Bar{t}}{c}\right)+\frac{v_x}{c}\cosh \left(\frac{a_x\,\Bar{t}}{c}\right)\Bigg]\label{be35}\\
   \text{and\,\,\,\,\,\,} \left(x+\gamma\frac{c^2}{a_x}\right)&=\gamma\left(\frac{c^2}{a_x}+\Bar{x}\right)\Bigg[\cosh \left(\frac{a_x\, \Bar{t}}{c}\right)+\frac{v_x}{c}\sinh \left(\frac{a_x\,\Bar{t}}{c}\right)\Bigg]\label{be36}
\end{align}
Now dividing \eqref{be35} by \eqref{be36} we obtain:
\begin{align}
    \frac{\left(ct+\gamma\frac{v_xc}{a_x}\right)}{\left(x+\gamma\frac{c^2}{a_x}\right)}&=\frac{\sinh \left(\frac{a_x\, \Bar{t}}{c}\right)+\frac{v_x}{c}\cosh \left(\frac{a_x\,\Bar{t}}{c}\right)}{\cosh \left(\frac{a_x\, \Bar{t}}{c}\right)+\frac{v_x}{c}\sinh \left(\frac{a_x\,\Bar{t}}{c}\right)}=\frac{\tanh \left(\frac{a_x\, \Bar{t}}{c}\right)+\frac{v_x}{c}}{1+\frac{v_x}{c}\cdot\tanh \left(\frac{a_x\, \Bar{t}}{c}\right)}\label{be37}
\end{align}
Then using the identity, $\tanh(A+B)=\frac{\tanh A+\tanh B}{1+\tanh A\tanh B}$ we obtain from above as:

\begin{align}
    &\frac{\left(ct+\gamma\frac{v_xc}{a_x}\right)}{\left(x+\gamma\frac{c^2}{a_x}\right)}=\frac{\tanh \left(\frac{a_x\, \Bar{t}}{c}\right)+\tanh\left\{\tanh^{-1}\left(\frac{v_x}{c}\right)\right\}}{1+\tanh \left(\frac{a_x\, \Bar{t}}{c}\right)\tanh\left\{\tanh^{-1}\left(\frac{v_x}{c}\right)\right\}}\nonumber\\
    &\hspace{2.4cm}=\tanh\Bigg\{\frac{a_x\, \Bar{t}}{c}+\tanh^{-1}\left(\frac{v_x}{c}\right)\Bigg\}\nonumber\\
    \text{or\,\,\,\,\,\,}&\frac{a_x\, \Bar{t}}{c}=\tanh^{-1}\Bigg\{\frac{\left(ct+\gamma\frac{v_xc}{a_x}\right)}{\left(x+\gamma\frac{c^2}{a_x}\right)}\Bigg\}-\tanh^{-1}\left(\frac{v_x}{c}\right)\label{be38}
\end{align}
Again using the identity, $\tanh^{-1}A-\tanh^{-1}B=\tanh^{-1}\Big\{\frac{A-B}{1-AB}\Big\}$ we obtain as:
\begin{align}
    c\Bar{t}=\frac{c^2}{a_x}\tanh^{-1}\left\{\frac{\left(ct+\gamma\frac{v_xc}{a_x}\right)-\frac{v_x}{c}\left(x+\gamma\frac{c^2}{a_x}\right)}{\left(x+\gamma\frac{c^2}{a_x}\right)-\frac{v_x}{c}\left(ct+\gamma\frac{v_xc}{a_x}\right) }\right\}\label{be39}
\end{align}
Additionally, squaring both sides of equations \eqref{be35} and \eqref{be36} we can obtain as follows:
\begin{align}
      \left(x+\gamma\frac{c^2}{a_x}\right)^2&-\left(ct+\gamma\frac{v_xc}{a_x}\right)^2=\gamma^2\left(\frac{c^2}{a_x}+\Bar{x}\right)^2\Bigg[\Bigg\{\cosh \left(\frac{a_x\, \Bar{t}}{c}\right)\nonumber\\
      &+\frac{v_x}{c}\sinh \left(\frac{a_x\,\Bar{t}}{c}\right)\Bigg\}^2-\Bigg\{\sinh \left(\frac{a_x\, \Bar{t}}{c}\right)+\frac{v_x}{c}\cosh \left(\frac{a_x\,\Bar{t}}{c}\right)\Bigg\}^2\Bigg]\nonumber\\
      &=\gamma^2\left(\frac{c^2}{a_x}+\Bar{x}\right)^2\Bigg[1-
      \frac{v_x^2}{c^2}\Bigg]=\left(\frac{c^2}{a_x}+\Bar{x}\right)^2\nonumber\\
      \text{or\,\,\,\,\,\,}\Bar{x}&=\Bigg[\left(x+\gamma\frac{c^2}{a_x}\right)^2-\left(ct+\gamma\frac{v_xc}{a_x}\right)^2\Bigg]^{1/2}-\frac{c^2}{a_x}\label{be40}
\end{align}
Therefore the inverse Rindler transformation for an accelerated frame having non-zero initial velocity can be written by using equations \eqref{be39} and \eqref{be40} as follows:
\begin{align}
c\Bar{t}&=\frac{c^2}{a_x}\tanh^{-1}\left\{\frac{\left(ct+\gamma\frac{v_xc}{a_x}\right)-\frac{v_x}{c}\left(x+\gamma\frac{c^2}{a_x}\right)}{\left(x+\gamma\frac{c^2}{a_x}\right)-\frac{v_x}{c}\left(ct+\gamma\frac{v_xc}{a_x}\right) }\right\}\\
    \Bar{x}&=\Bigg[\left(x+\gamma\frac{c^2}{a_x}\right)^2-\left(ct+\gamma\frac{v_xc}{a_x}\right)^2\Bigg]^{1/2}-\frac{c^2}{a_x}\\
    \Bar{y}&=y\\
    \Bar{z}&=z
\end{align}

\section*{Acknowledgments}
We thank faculty members and colleagues of the Department of Physics, Assam University, Silchar-788011 (India) for their encouragement and support while doing this work.

%\bibliographystyle{unsrt}
%\bibliography{Brahma_Sen_m}

\end{document}